\documentclass[prd,10pt,twoside,a4paper,tightenlines,nofootinbib]{revtex4}
\usepackage{epsf,graphicx,mathrsfs,amsmath,amssymb,amsthm,amsopn,mathbbol,bm,feynmf}
\begin{document}
\newcommand{\bc}{\begin{center}}
\newcommand{\ec}{\end{center}}
\newcommand{\be}{\begin{equation}}
\newcommand{\ee}{\end{equation}}
\newcommand{\bea}{\begin{eqnarray}}
\newcommand{\eea}{\end{eqnarray}}
\newcommand{\bi}{\begin{itemize}}
\newcommand{\ei}{\end{itemize}}
\newcommand{\ts}[1]{\mbox{\scriptsize {#1}}}
\newcommand{\tst}[1]{\mbox{\tiny {#1}}}
\newcommand{\meanf}{\phi}
\newcommand{\prop}{G}
\newcommand{\cint}{\int_{\mathcal{C}}}
\newcommand{\half}{\frac{1}{2}}
\newcommand{\qbrs}{Q_{\tst{\sc BRS}}}
\newcommand{\phix}{\varphi}

\title{Gauge-fixing dependence of $\Phi$-derivable approximations}
\author{A. Arrizabalaga, J. Smit} 
\affiliation{Institute for Theoretical Physics, University of Amsterdam,
Valckenierstraat 65, 1018 XE Amsterdam, The Netherlands}
\date{\today}

\begin{abstract}
{\bf Abstract.} We examine the problem of gauge dependence of the
2PI effective action and its $\Phi$-derivable approximations in gauge
theories. The dependence on the gauge-fixing condition is obtained. The result
shows that $\Phi$-derivable approximations, defined as truncations of the 2PI effective
action at a certain order, have a controlled gauge dependence,
i.e. the gauge dependent terms appear at higher order than the truncation
order. Furthermore, using the stationary point obtained for the approximation to
evaluate the complete 2PI effective action boosts the order at which the gauge
dependent terms appear to twice the order of truncation. We also comment on
the significance of this controlled gauge dependence.
\end{abstract}

\maketitle

\section{Introduction}

Perturbative approaches to the study of equilibrium and non-equilibrium properties of hot and dense media may lead to inconsistencies and are often plagued with infrared divergences. These problems are linked to the fact that calculations in terms of the bare quantities of the underlying quantum field theory (and perturbative approximations thereof) fail to describe the collective phenomena in the medium. A strategy to tackle this handicap of the theory is to work with dressed quantities, in which the most relevant effects of the interacting ensemble are accounted for. These dressed quantities are obtained by means of non-perturbative resummation schemes, which usually involve solving a set of self-consistent equations.\\

An arbitrary resummation scheme will however not guarantee that the
conservation laws of the original theory are preserved by the dressed
quantities. A way to solve this problem is by formulating the scheme in terms
of an action functional that respects the symmetries of the original theory. A
particular kind of such action functionals was first introduced
in the study of non-relativistic Fermi systems by Luttinger and
Ward~\cite{Luttinger:1960}, De Dominicis and Martin~\cite{DeDominicis} and
Baym~\cite{Baym:1962} and later generalized to relativistic field theories by
Cornwall, Jackiw and Toumboulis~\cite{Cornwall:1974vz}. These functionals,
which are derived
from the so-called 2PI effective action, involve a diagrammatic expansion in
terms of
two-particle irreducible (2PI) skeleton graphs.
A particular choice for an action functional is obtained by
truncating this diagrammatic series. This defines what is called a
$\Phi$-derivable approximation. A variational principle applied to the
resulting action leads to a set of self-consistent equations from which the dressed quantities are obtained.\\

A manifest advantage of such a functional formulation is that global
symmetries of the original theory are preserved. Additionally, the variational
principle used to determine the dressed quantities guarantees thermodynamic
consistency \cite{Baym:1962}.  All these useful properties make $\Phi$-derivable
approximations a very attractive mathematical framework for the
study of properties of high-energy plasmas. In particular, they may prove
useful for QCD plasmas, whose interest has grown in recent years due to the
possibility of creating quark-gluon plasma in heavy-ion collision experiments
at Brookhaven and CERN. Calculations of thermodynamical quantities such as the
entropy~\cite{Blaizot:2000fc} and free energy~\cite{Peshier:2000hx} have been
achieved using these methods. In those calculations a resummation of the
physics encoded in the hard thermal loops (HTL) was performed. Important to
mention is the fact that, due to the remarkable symmetry properties of the
HTL, the results in Refs.~\cite{Blaizot:2000fc,Peshier:2000hx} are manifestly gauge
invariant. Non-equilibrium properties can also be formally
studied within these approximation schemes~\cite{Ivanov:1998nv}. They could be
used to shed some light on important issues such as thermalization and loss of initial correlations. Very interesting results in this direction have been obtained~\cite{Berges:2000ur} with scalar models. \\

However, an extension of these approximation schemes beyond the HTL regime in
the study of QCD plasmas is still lacking. There are two main problems
involved. One is that renormalization seems to be a non-trivial issue, as
shown in explicit calculations for scalar theories \cite{Braaten:2001vr}. To
deal with this obstacle, a recent approach based on BPHZ renormalization has
been proposed by van Hees and Knoll~\cite{vanHees:2001ik}. 
The other main problem is the fact that gauge invariance may be lost in the
approximations. This is because, in general, the solutions for the dressed
propagators and/or vertices do not satisfy Ward identities. In particular this implies that thermodynamical quantities computed within these approximations will suffer from gauge dependence. This pathology shows up as an explicit dependence on the choice of gauge-condition.\\

In this paper we study the problem of gauge dependence of the 2PI effective action and
its $\Phi$-derivable approximations. In Sec.~\ref{sec:two} we review the
general formalism of $\Phi$-derivable approximations and introduce the
notation to be used. In
Sec.~\ref{sec:three} we apply the formalism to gauge theories and
determine the dependence of the 2PI effective action under a change of the
gauge-fixing condition. From the result one sees that the 2PI effective action
is gauge independent at its stationary point. This was already shown for the
1PI effective action and expected from general
arguments~\cite{Nielsen:1975fs}. In Sec.~\ref{sec:four} we
apply the result of Sec.~\ref{sec:three} to the $\Phi$-derivable
approximations that result from truncating the 2PI effective action at a
certain order. We show that these approximations have a controlled
gauge-fixing dependence, i.e. the gauge-dependent terms appear at higher
order. We discuss in Sec.~\ref{sec:five} that the use of $\Phi$-derivable
approximations restricts the choices of gauge fixing available, if they are
indeed to be good approximations to the exact theory. This prevents the
high-order gauge-dependent terms to take arbitrarily large values, such that the gauge
dependence will be indeed controlled. 

\section{2PI effective action and $\Phi$-derivable approximations}\label{sec:two}
The generating functional for correlation functions can be written as
\begin{equation}
Z[J,K]=\int \mathcal{D}\varphi\, e^{i\left\{S[\varphi]+J_i\varphi^i+\half \varphi^i K_{ij} \varphi^j\right\}},
\label{Z2}
\end{equation}
where $S[\phix]$ is the action, $\phix$ represents the fields and the $J$ and $K$ are auxiliary external sources. We use a shorthand notation where Latin indices stand for all field and current attributes (i.e. $\phix(x) \rightarrow \phix_i$) and summation and/or integration over repeated indices is understood, i.e. $J_i\phix^i=\int d^4x\, J(x)\phix(x)$.\footnote{The time integration involved in this functional product can also run along a contour $\mathcal{C}$ in the complex plane such as the ones used in the real and imaginary time formalisms of thermal field theory. This detail will however not be important in our calculations, so we will omit the subscript $\mathcal{C}$ in the integrations.}. The generating functional of connected diagrams $W$ is defined from $Z$ as
\begin{equation}
W[J,K]=-i\log \left( Z[J,K]\right).
\end{equation}
The expectation value of a functional $O[\phix]$ is given by
\
\begin{equation}
\left\langle O[\phix] \right\rangle\equiv\frac{\int \mathcal{D}\phix\, O[\phix]\,e^{i\left\{S[\phix]+J_i\phix^i+\frac{1}{2}\phix^i  K_{ij}\phix^j\right\}}}{\int \mathcal{D}\phix\,e^{i\left\{S[\phix]+J_i\phix^i+\frac{1}{2}\phix^i  K_{ij}\phix^j\right\} }}=iO\left[\frac{\delta}{\delta (iJ)}\right]W.
\label{expectationvalues}
\end{equation}
Mean fields $\meanf^i$ and connected correlation functions $G^{ijk\ldots}$ can then be obtained by functional differentiations of $W[J]$ as
\begin{equation}
i\frac{\delta W}{\delta (iJ_i)}=\meanf^i, \ \ \ i\frac{\delta^2 W}{\delta (iJ_i) \delta (iJ_j)}=G^{ij}, \ \ \ i\frac{\delta^N W}{\delta (iJ_{i}) \delta (iJ_j) \delta (iJ_k) \ldots }=G^{ijk\ldots}.
\end{equation}

Functional differentiations of $W[J,K]$ with respect to the bilocal currents $K$ may generate also disconnected diagrams. For example, differentiating once with respect to $K$ leads to
\begin{equation}
i\frac{\delta W[J,K]}{\delta (iK_{ij})}=\half\left( \meanf^{i}\meanf^{j} + \prop^{ij} \right).
\label{functionalw}
\end{equation}
A functional Legendre transform in the mean field $\meanf^i$ and the two-point
function $\prop^{ij}$ leads to the so-called 2PI effective action
\begin{equation}
\Gamma[\meanf,\prop]=W[J,K]-J_i\meanf^i-\half K_{ij} \left(\meanf^i\meanf^j+\prop^{ij}\right),
\label{legendre}
\end{equation}
From its definition one can derive the relations
\begin{equation}
\frac{\delta \Gamma[\meanf,\prop]}{\delta \meanf^i}=-J_i-K_{ij}\meanf^j\ \ {\rm{and}} \ \ \frac{\delta \Gamma[\meanf, \prop]}{\delta \prop^{ij}}=-\half K_{ij}.
\label{trafos}
\end{equation}
With the help of Eq.~(\ref{trafos}) one can write the expression for the expectation value of a functional $O[\phix]$ in terms of the 2PI effective action as
\begin{equation}
\langle O[\phix]\rangle=e^{-i\Gamma[\meanf,G]}\int \mathcal{D}\phix\,O[\phix]\, e^{i\left\{ S[\phix]-\frac{\delta \Gamma[\meanf,G]}{\delta \meanf_i}(\phix-\meanf)_i-\frac{\delta \Gamma[\meanf,G]}{\delta G_{ij}}\left[ (\phix-\meanf)_i(\phix-\meanf)_j-G_{ij}\right]\right\}}.
\label{expectationvalueswitheffectiveaction}
\end{equation}
The 2PI effective action can be cast into the very convenient form~\cite{Luttinger:1960,DeDominicis,Baym:1962,Cornwall:1974vz,Ivanov:1998nv} 
\begin{equation}
\Gamma[\meanf,\prop]=S_{0}[\meanf]+ic\,\rm{Tr}\left\{ \log\left(
\prop^{-1}\right)+\prop \left(\prop_{0}^{-1}-\prop^{-1} \right) \right\}-i\Phi[\meanf,\prop].
\label{convenientform}
\end{equation} 
where $S_0$ is the free part of action, $\prop_{0}$ is the bare two-point function
$\left(-i\delta^2 S_0[\phix]/\delta \phix \delta \phix\right)^{-1}$  and $c$
is a constant equal to $1/2$ for bosons and $-1$ for fermions. The functional $\Phi[\meanf, G]$ consists on the sum of all
two-particle-irreducible (2PI) \emph{skeleton diagrams} with bare vertices and
dressed propagators. In this context skeleton diagrams are those without
self-energy insertions. Non-2PI diagrams with mean field
insertions are also included in the definition of $\Phi$.\footnote{In the literature~\cite{Cornwall:1974vz} $\Phi$ is usually
defined in such a way that it only contains strict 2PI diagrams. This involves a redefinition of the
action to include mean fields and tadpoles. We prefer the above notation where
all interaction parts are placed in $\Phi$. Of course, both definitions agree
when $\meanf=0$.} For example, in the case of a theory with quartic
interactions (such as $\lambda \phi^4$), and using Dyson relation $\prop^{-1}=\prop_{0}^{-1}+i\Pi$ between the two-point function and the self-energy $\Pi$ , the above expression can be written graphically as
 \begin{equation}
\Gamma[\meanf,\prop]=S_0[\meanf]+\parbox[c]{12cm}{
\epsfxsize=12cm
\epsfbox{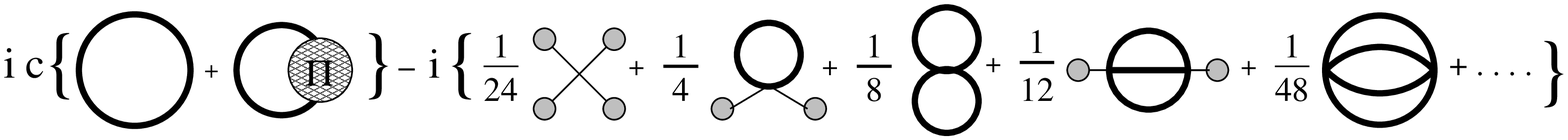}}
\label{skeletonloopexpansion}
\end{equation}
where the thick lines are \emph{dressed} propagators, the small lollipops are
the mean fields and the cross-hatched blob is the one-particle irreducible
self-energy $\Pi$. Writing the effective action in terms of skeleton diagrams
makes possible to incorporate higher-order effects into the propagators with
correct prefactors. \\ 

In practice one has to restrict to an approximated version of the
2PI effective action that results from considering only a certain subset of
diagrams in the functional $\Phi$. This defines a $\Phi$-derivable
approximation. One typically considers the loop expansion of skeleton diagrams
as pictured in Eq.~(\ref{skeletonloopexpansion}) (and refered to as the
\emph{skeleton-loop expansion} in the following), and truncates it at a given
order. In this way the 2PI effective action $\Gamma$ is split into
two pieces: the truncated part $\Gamma_0$, and the higher order part
$\Gamma_1$. Then one takes $\Gamma_0$ as the approximated effective action.
This action defines \emph{approximate} mean fields and two-point functions
$\meanf_{\ts{ap}}$ and $\prop_{\ts{ap}}$, which result from the stationarity
condition, i.e. from the implicit functional equation (\ref{trafos}) for vanishing sources $J$ and $K$, as

\begin{equation}
\frac{\delta \Gamma_0[\meanf,G]}{\delta \meanf}\Big|_{\meanf_{\ts{ap}} ,\,
\prop_{\ts{ap}} }=0  \ \ {\rm{and}}\ \ \frac{\delta \Gamma_0[\meanf,G]}{\delta \prop}\Big|_{\meanf_{\ts{ap}} ,\,
\prop_{\ts{ap}}}=0.
\label{stationaryrequirementtruncated}
\end{equation} 
These \emph{approximate} mean fields and two-point functions defined from the truncated action $\Gamma_0$ differ from the \emph{exact} ones, which are obtained from the stationary point of the complete 2PI effective action as
\begin{equation}
\frac{\delta \Gamma[\meanf,G]}{\delta \meanf}=\frac{\delta (\Gamma_0+\Gamma_1)}{\delta \meanf}\Big|_{\meanf_{\ts{ex}} ,\,
\prop_{\ts{ex}} }=0  \ \ {\rm{and}}\ \ \frac{\delta \Gamma[\meanf,G]}{\delta \prop}=\frac{\delta (\Gamma_0+\Gamma_1)}{\delta \prop} \Big|_{\meanf_{\ts{ex}} ,\, \prop_{\ts{ex}}}=0.
\label{stationaryrequirement} 
\end{equation} 
We end this section by noting that one could also construct more general effective actions by including higher-point external sources $J_{i}$,$K_{ij}$,$L_{ijk} \ldots$ into the functional $W=W[J_i,K_{ij},L_{ijk},\ldots]$ and performing a Legendre transform as follows
\bea
\Gamma[\meanf_i,\prop_{ij},G_{ijk},\ldots]&=&W[J_i,K_{ij},L_{ijk},\ldots]\nonumber \\
&&-J_i\meanf^i-\half K_{ij}\left( \meanf^{i}\meanf^{j}+\prop^{ij}\right)-\frac{1}{6}L_{ijk}\left(G^{ijk}+3\prop^{ij}\meanf^{k}+\meanf^{i}\meanf^{j}\meanf^{k} \right)-\ldots.
\eea
This form of the effective action can be rewritten as a diagrammatic series in terms of skeleton diagrams of the $n$-point vertex functions~\cite{Norton:1975bm} and can be used for generalized $\Phi$-derivable approximations.
\section{Gauge-fixing dependence of the 2PI effective action}\label{sec:three}
We consider now the case of a pure Yang-Mills theory with gauge group $G$. Its action is given by
\begin{equation}
S_{\ts{YM}}=-\int d^{4} x \, \frac{1}{4}F_{\mu \nu}^{a}(x)F^{\mu \nu}_{a}(x),
\label{gaugeaction}
\end{equation}
where $F_{\mu \nu}\equiv F_{\mu
\nu}^{a}\,T_a=\partial_{\mu}A_{\nu}-\partial_{\nu}A_{\mu}-g\left[A_{\mu},A_{\nu}\right]$
is the field-strength tensor of the gauge field $A_{\mu}=A_{\mu}^a T_a$, $g$
is the (unrenormalized) coupling constant and $T_a$ are the generators of the Lie algebra of the gauge group $G$.\\

The action is invariant under gauge transformations $U(x)\in G$ of the gauge potential $A_{\mu}$
\begin{equation}
A_{\mu} \rightarrow\ \  {^{U}}\!\!A_{\mu} (x)=U(x)A_{\mu}(x)U^{-1}(x)-\frac{i}{g} \left[ \partial_{\mu}U(x)\right] U^{-1}(x).
\label{gaugetrafos}
\end{equation}
This invariance implies that the functional integrals over gauge field
configurations are ill-defined. One gets around this difficulty by the
Faddeev-Popov gauge-fixing procedure, which introduces a gauge-breaking term
$S_{\ts{GF}}$ into the action. In the context of BRS-quantization this term is realized in a useful manner by introducing some auxiliary fields: the Faddeev-Popov fermionic ghost fields $c^a$ and $\bar{c}^a$ and the bosonic Lautrup-Nakanishi fields $B^a$. The gauge-fixing is implemented through the condition $C^a[A]=0$, where a typical choice is the covariant gauge $C^a[A]=\partial^{\mu}A^{a}_{\mu}$. The gauge-fixed action then reads
\begin{equation}
S=S_{\ts{YM}}+S_{\ts{GF}}=\int d^4 x\,\left\{-\frac{1}{4} F_{\mu
\nu}^{a}(x)F^{\mu \nu}_{a}(x)-\bar{c}_a(x) \frac{\delta C^a[A]}{\delta
A_{b\mu}}{\bf \big(}D_{\mu}c(x)\,{\bf \big)}_{b}+B_a(x)C^a[A]-\half \xi B_a(x)B^a(x) \right\},
\label{gaugefixedaction}
\end{equation}
where $D_{\mu}\equiv\partial_{\mu}-igT^{a}A_{a}$ is the covariant derivative and $\xi$ is the gauge-fixing parameter.
\\

The action obtained by adding this gauge-fixing term is no longer invariant under local gauge transformations (\ref{gaugetrafos}). However, it is invariant under BRS transformations, which are defined as 
\bea
\delta_{\tst{BRS}}A_{\mu}^{a}&=&\epsilon\left( D_{\mu}c\right)^{a}, \nonumber \\
 \delta_{\tst{BRS}}c^a&=&i\epsilon g \, c^2,  \nonumber \\
\delta_{\tst{BRS}}\bar{c}^{a}&=&-\epsilon  B^{a}, \nonumber \\
 \delta_{\tst{BRS}}B^a&=&0, 
\label{BRStrafos}
\eea    
where $\epsilon$ is an infinitesimal global anti-commuting parameter and $c^2$ is a short-hand notation for $(T_ac^a)(T_bc^b)=1/2 [T^a,T^b]c^ac^b$. The Lautrup-Nakanishi field $B$ has been introduced to ensure the nilpotency of the BRS charge $Q_{\tst{BRS}}$, defined as $\delta_{\tst{BRS}}=\epsilon  Q_{\tst{BRS}}$. It allows for a convenient rewriting of the gauge-breaking term as a complete BRS variation 
\begin{equation}
S_{\tst{\sc GF}}=Q_{\tst{\sc BRS}}\int d^4x\,  \left\{\half \,\xi \bar{c}_a(x) B^a(x)-\bar{c}_a(x)C^a[A]\right\}\equiv Q_{\tst{\sc BRS}} \Psi.
\label{definitionofpsi}
\end{equation} 
Using the notation of Sec.~\ref{sec:two} the generating functional $Z[J,K]$ for this gauge theory can be compactly written as
\begin{equation}
Z[J,K]=\mathcal{N}_{\xi}\int \mathcal{D} \phix \, e^{i\left\{ S_{\tst{\sc
YM}}+Q_{\tst{\sc BRS}}\Psi +J_i \phix^i+\half \phix^i K_{ij} \phix^j \right\}
}= e^{i\left\{\Gamma[\meanf,\prop]+J_i \meanf^i+\half K_{ij} \left(\meanf^i \meanf^j +\prop^{ij} \right)\right\}},
\label{efwithw}
\end{equation}

where $\phix$ denotes collectively all fields $\left\{ A_{\mu}^{a}(x),
c^{a}(x), \bar{c}^{a}(x), B^a(x) \right\}$, $J$ and $K$ denote all their
associated currents $\left\{J_A,J_c,J_{\bar{c}},J_B,K_{AA},K_{c\bar{c}},K_{BB}
\right\}$, and Latin indices stand for both space-time and group indices, i.e.
$A_{\mu}^{a}(x) \rightarrow A^{i}$. This notation is used to allow one to
write formulas in a compact way, though one should bear in mind that the ghost fields $c$ and $\bar{c}$ and their associated
local currents $J_{c}$ and $J_{\bar{c}}$ are anti-commuting variables.
$\mathcal{N}_{\xi}$ is a $\xi$-dependent infinite constant generated during
the Faddeev-Popov gauge-fixing procedure. Its gauge parameter dependence can
be seen already in the free theory and can be absorbed into the action by rescaling the ghost fields by $\xi^{-1/4}$.
Hence this constant will not play a role in the following.  \\

Having set up the notation we turn now to study the gauge dependence of the
2PI effective action. We study how it transforms both under a change of the gauge-fixing condition $C^a[A]\rightarrow C^a[A]+\Delta C^a[A]$ and gauge parameter $\xi \rightarrow \xi+\Delta \xi$, or more generally, under a change $\Psi \rightarrow \Psi+\Delta \Psi$. Under this shift of gauge condition, the effective action, the mean field and the two-point function respectively change as
\begin{eqnarray}
\Gamma \rightarrow \Gamma'=\Gamma+\Delta \Gamma\ ,\ \
 \meanf \rightarrow \meanf'=\meanf+\Delta \meanf \ {\rm{and}} \ \ 
\prop \rightarrow \prop'=\prop+\Delta  \prop. 
\label{changegauge}
\end{eqnarray}
The currents $J_i$ and $K_{ij}$ are taken to be gauge independent since they
are external. This fact allows us to calculate immediately from
Eq.~(\ref{trafos}) how much the first functional derivatives of the effective
action vary under the gauge-fixing change, obtaining 
\begin{equation}
\Delta \left( \frac{\delta \Gamma}{\delta \prop_{ij}} \right)=0 \ \ {\rm{and}} \ \ 
\Delta \left( \frac{\delta \Gamma}{\delta \meanf_i} \right)=2\Delta \meanf_j  \frac{\delta \Gamma}{\delta \prop_{ij}}.
\label{changeofcondition}
\end{equation}
The first functional derivatives of $\Gamma$ are used to find the stationary
point by setting them to zero as done in Eq.~(\ref{stationaryrequirement}).
From Eq.~(\ref{changeofcondition}) one notices that the stationarity condition itself is only gauge invariant if it is realized \emph{simultaneously} for both arguments $\meanf$ and $\prop$.\\

To compute the variation of the effective action $\Gamma$ itself one can use
the relations (\ref{trafos}) to cast Eq.~(\ref{efwithw}) into the convenient form
\begin{equation}
e^{i\Gamma_{\Psi}[\meanf,\prop]}=\int \mathcal{D} \phix \, e^{i\left\{S_{\tst{\sc YM}}+Q_{\tst{\sc BRS}}\Psi - (\phix-\meanf)_{i} \frac{\delta \Gamma}{\delta \meanf_{i}} - \left[ (\phix_{i}-\meanf_i)(\phix_{j}-\meanf_j)-\prop_{ij}\right]\frac{\delta \Gamma}{\delta \prop_{ij}}  \right\} }.
\label{beforegaugefixing}
\end{equation}
For simplicity and later convenience we denote the field combinations
$(\phix-\meanf)_i$ and
$\left[(\phix-\meanf)_i(\phix-\meanf)_j-\prop_{ij}\right]$ that appear in the
exponent of Eq.~(\ref{beforegaugefixing}) as $\widetilde{\phix}_i$ and
$\widetilde{G}_{ij}$ respectively. These have the property that their
expectation values $\langle \widetilde{\phix}_i \rangle$ and $\langle \widetilde{\prop}_{ij}\rangle$ vanish. \\

After a change of gauge condition $\Psi \rightarrow \Psi'=\Psi+\Delta \Psi$ equation (\ref{beforegaugefixing}) becomes
\bea
e^{i\Gamma'}=e^{i\left( \Gamma+\Delta \Gamma\right)}=\int \mathcal{D} \phix \,
\biggl\{e^{i\left\{S_{\tst{\sc YM}}+Q_{\tst{\sc BRS}}\Psi - \widetilde{\phix}_{i}\frac{\delta
\Gamma}{\delta \meanf_{i}} -\widetilde{G}_{ij}\frac{\delta \Gamma}{\delta
\prop_{ij}}\right\}} \,\,e^{i\left\{\qbrs \Delta \Psi+\Delta \meanf_i\frac{\delta \Gamma}{\delta \meanf_i} + \Delta \prop_{ij}\frac{\delta \Gamma}{\delta \prop_{ij}}+\Delta \meanf_i \Delta \meanf_j\frac{\delta \Gamma}{\delta \prop_{ij}} \right\}} \biggr\},
\eea
which using the notation of Eq.~(\ref{expectationvalues}) leads to 
\begin{equation}
e^{i\Delta \Gamma}= \left\langle e^{i\left\{\qbrs \Delta \Psi+\Delta \meanf_i\frac{\delta \Gamma}{\delta \meanf_i} + \Delta \prop_{ij}\frac{\delta \Gamma}{\delta \prop_{ij}}+\Delta \meanf_i \Delta \meanf_j\frac{\delta \Gamma}{\delta \prop_{ij}} \right\}} \right\rangle \label{change}.
\end{equation}
This result is valid for any finite change in the gauge-fixing conditions. To
proceed further we restrict ourselves to infinitesimal variations $\Delta
\Psi$. Then one can expand both sides of Eq.~(\ref{change}) to obtain 
\bea
\Delta \Gamma = \left\langle \qbrs \Delta \Psi \right\rangle + \Delta \meanf_{i}\frac{\delta \Gamma}{\delta \meanf_i}+\Delta \prop_{ij}\frac{\delta \Gamma}{\delta \prop_{ij}}+O(\Delta^2) \label{eq:preexpandedchange}
\eea
where we used the fact that $\Delta \meanf$ and $\Delta \prop$ are of order
$O(\Delta \Psi)$. This can be easily checked. Indeed, following the same steps
as to obtain Eq.~(\ref{change}) one gets for the mean field
\begin{equation}
\meanf'=\meanf+ \Delta\meanf =e^{-i\Delta\Gamma} \left\langle \phix\, e^{i\left\{ \qbrs \Delta \Psi+\Delta \meanf_i\frac{\delta \Gamma}{\delta \meanf_i} + \Delta \prop_{ij}\frac{\delta \Gamma}{\delta \prop_{ij}}+\Delta \meanf_i \Delta \meanf_j\frac{\delta \Gamma}{\delta \prop_{ij}} \right\}} \right\rangle \label{meanchange},
\end{equation}
which using Eq.~(\ref{change}) reduces to 
\begin{equation}
\meanf+\Delta \meanf=\frac{\left\langle \phix\, e^{i\qbrs \Delta \Psi}\right\rangle}{\left\langle e^{i\qbrs \Delta \Psi}\right\rangle},
\end{equation}
and expanding in $\Delta \Psi$ yields
\begin{equation}
\Delta \meanf_{i}=i\left\langle \widetilde{\phix}_i \qbrs \Delta \Psi \right\rangle + O(\Delta^2).
\label{eq:meanfieldvariation}
\end{equation}
Similarly one computes the variation of the two-point function obtaining
\begin{equation}
\Delta \prop_{ij}=i\left\langle \widetilde{G}_{ij} \qbrs \Delta \Psi \right\rangle +O(\Delta^2),
\label{eq:propagatorvariation}
\end{equation}
which verify our statement above. Moreover, Eqs.~(\ref{eq:meanfieldvariation})
and (\ref{eq:propagatorvariation}) can be used to write Eq.~(\ref{eq:preexpandedchange}) as
\begin{equation}
\Delta \Gamma=\left\langle \qbrs \Delta \Psi \right\rangle +i\left\langle \widetilde{\phix}_i \qbrs \Delta \Psi \right\rangle\frac{\delta \Gamma}{\delta \meanf_i} +i\left\langle \widetilde{G}_{ij}\qbrs \Delta \Psi \right\rangle\frac{\delta \Gamma}{\delta \prop_{ij}} +O(\Delta^2).
\label{eq:expandedchange}
\end{equation}
One expects the stationary point of the effective action, i.e. when its
functional derivatives are set to zero, to be gauge-independent. That is still
not obvious from Eq.~(\ref{eq:expandedchange}) since it appears that the first term in the r.h.s. would not vanish. For that, one can make use of the following trick~\cite{Iguri:2001za}. Consider the expectation value of the gauge-fixing change $\Delta \Psi$, namely
\begin{equation}
\langle \Delta \Psi \rangle=e^{-i\Gamma}\int \mathcal{D}\phix\,  \Delta \Psi\, e^{i\left\{S_{\tst{\sc YM}}+Q_{\tst{\sc BRS}}\Psi - \widetilde{\phix}_{i}\frac{\delta \Gamma}{\delta \meanf_{i}} -\widetilde{G}_{ij}\frac{\delta \Gamma}{\delta \prop_{ij}}\right\}}.
\label{twentynine}
\end{equation}
One can do a BRS transformation $\phix \rightarrow  \phix+\epsilon\qbrs \phix$ on the field variables in the path-integral. This transformation leaves the measure invariant and only amounts to a shift of the integration variable, so the equation remains the same. The l.h.s can be however rewritten so that 
\begin{equation}
\langle \Delta \Psi \rangle= e^{-i\Gamma}\,\int \mathcal{D}\phix\, \left(
\Delta \Psi+\epsilon \qbrs \Delta \Psi \right) e^{i\left\{S_{\tst{\sc
YM}}+Q_{\tst{\sc BRS}}\Psi - 
\widetilde{\phix}_{i}\frac{\delta \Gamma}{\delta \meanf_{i}} -\widetilde{G}_{ij}\frac{\delta \Gamma}{\delta
\prop_{ij}}-\epsilon \,\qbrs \widetilde{\phix}_i\frac{\delta \Gamma}{\delta
\meanf_i} -\epsilon \,\qbrs \widetilde{G}_{ij}\frac{\delta \Gamma}{\delta
\prop_{ij}}\right\}},
\label{eq:brsshift}
\end{equation}
where we have used the fact that $\Delta \Psi[\phix+\epsilon\qbrs
\phix]=\Delta \Psi+\epsilon \qbrs \Delta \Psi$, which follows from the
definition (\ref{definitionofpsi}). The BRS charge $Q_{\tst{\sc BRS}}$ appearing in this expression does not operate on the mean fields $\meanf$ and two-point functions $\prop$ that are part of $\widetilde{\phix}$ and $\widetilde{G}$, but only on the fields to be path-integrated over. 
Expanding the r.h.s. of (\ref{eq:brsshift}) in the anticommuting parameter $\epsilon$ leads
to
\begin{equation}
\left\langle \qbrs \Delta \Psi \right\rangle=-i \left\langle \Delta \Psi \,\qbrs \left( \widetilde{\phix}_i\frac{\delta \Gamma}{\delta
\meanf_i}\right) \right\rangle-i \left\langle  \Delta \Psi \,\qbrs
\left(\widetilde{G}_{ij}\frac{\delta \Gamma}{\delta \prop_{ij}}\right)\right\rangle.
\label{eq:yepzero}
\end{equation} 
where the quantities have been reorganized so that the equation is valid
for all fields, both commuting ($A_{\mu}$ and $B$) and anticommuting
($\bar{c}$ and $c$). Combinations like
$\widetilde{\phix}\,\delta\Gamma/\delta \meanf$ or
$\widetilde{G}\,\delta\Gamma/\delta G$ are always commuting so it is
preferable to have them in this form.\\

The same procedure can be applied also to the expectation values $\langle
\widetilde{\phix}_a \Delta \Psi\rangle$ and $\langle\widetilde{G}_{ab} \Delta \Psi\rangle $ to obtain
\bea
\left\langle \widetilde{\phix}_{a}\, \qbrs \Delta \Psi
\right\rangle&=&\left\langle\Delta \Psi  \,\qbrs\widetilde{\phix}_{a}
\right\rangle-i \left\langle \widetilde{\phix}_{a}\Delta
  \Psi \,\qbrs\left(\widetilde{\phix}_j\frac{\delta
\Gamma}{\delta \meanf_j}\right)  \right\rangle-i \left\langle  \widetilde{\phix}_{a}
  \Delta \Psi\,\qbrs\left(
  \widetilde{\prop}_{jk}\frac{\delta \Gamma}{\delta \prop_{jk}}\right)\right\rangle,\label{eq:yep1}\\
\left\langle \widetilde{G}_{ab}\, \qbrs \Delta \Psi
\right\rangle&=&\left\langle\Delta \Psi \,
\qbrs\widetilde{G}_{ab} \right\rangle-i \left\langle 
\widetilde{G}_{ab}\Delta \Psi \,\qbrs\left(\widetilde{\phix}_j\frac{\delta
\Gamma}{\delta \meanf_j}\right)  \right\rangle-i \left\langle \widetilde{G}_{ab}\,
  \Delta \Psi \,\qbrs\left(
  \widetilde{\prop}_{jk}\frac{\delta \Gamma}{\delta \prop_{jk}}\right) \right\rangle.
\label{eq:yep2}
\eea      
These results enable us to write the change in the effective action $\Delta
\Gamma$ only in terms proportional to its functional derivatives simply by substituting Eqs.~(\ref{eq:yepzero})-(\ref{eq:yep2}) into Eq.~(\ref{eq:expandedchange}). One notices that the first terms coming from the
r.h.s. of Eqs.~(\ref{eq:yep1}) and (\ref{eq:yep2}) cancel exactly those that
come from Eq.~(\ref{eq:yepzero}) when both are substituted into Eq.~(\ref{eq:expandedchange}).
In this way terms with a single functional derivative do not appear in $\Delta \Gamma$.
After some rearrangements one is then left with
\begin{equation}
\Delta \Gamma=\half \left\langle \Delta \Psi\, \qbrs
\left(\tilde{\phix}_i\frac{\delta \Gamma}{\delta \meanf_i}\,\tilde{\phix}_j\frac{\delta
\Gamma}{\delta \meanf_j} \right)\right\rangle+\left\langle \Delta \Psi\,
\qbrs \left(\tilde{\phix}_i\frac{\delta \Gamma}{\delta
\meanf_i}\,\widetilde{G}_{jk}\frac{\delta \Gamma}{\delta \prop_{jk}}
\right)\right\rangle+\half \left\langle \Delta \Psi\, \qbrs
\left(\widetilde{G}_{ij}\frac{\delta \Gamma}{\delta
\prop_{ij}}\,\widetilde{G}_{kl}\frac{\delta \Gamma}{\delta \prop_{kl}}\right)\right\rangle+O(\Delta^2),
\end{equation}
which can be cast into the compact result
\begin{equation}
 \Delta \Gamma \left[ \meanf, \prop\right]=\half \left\langle \Delta \Psi\, \qbrs \left( \widetilde{\phix}_i\frac{\delta \Gamma}{\delta \meanf_i} +\widetilde{G}_{jk}\frac{\delta \Gamma}{\delta \prop_{jk}} \right)^2 \right\rangle+O(\Delta^2)
\label{eq:fr},
\end{equation}
where the average and $\qbrs$ only apply to the fields $\phix$ contained in
$\widetilde{\phix}$ and $\widetilde{G}$.\\

Eq.~(\ref{eq:fr}) gives the variation of the 2PI effective action caused
by a change in the gauge condition and is the main result of this paper. One sees that when the functional derivatives of $\Gamma$ are set to zero this variation vanishes, and then the effective action is gauge-fixing independent. This situation occurs precisely at the stationary point, i.e. at the \emph{exact} mean fields $\meanf_{\ts{ex}}$ and two-point functions $\prop_{\ts{ex}}$.  \\

However, the quantities $\meanf_{\ts{ex}}$ and $\prop_{\ts{ex}}$ are not
gauge-fixing independent themselves. Indeed, one can explicitly compute their
gauge dependence by applying the condition (\ref{stationaryrequirement}) to
Eqs.~(\ref{eq:meanfieldvariation}) and (\ref{eq:propagatorvariation}), obtaining in this manner 
\bea
\Delta \meanf_{\ts{ex}}^{i}&=&i\left\langle \Delta \Psi \qbrs \left( \widetilde{\phix}_{\ts{ex}}\right)^i\right\rangle ,\\
\Delta \prop_{\ts{ex}}^{ij}&=&i\left\langle \Delta \Psi\qbrs \left( \widetilde{G}_{\ts{ex}}\right)^{ij} \right\rangle. 
\eea
For the case of the effective action $\Gamma [
\meanf_i,\prop_{ij},G_{ijk},\ldots]$ including higher-point correlation
functions, the same procedure leads to the generalized result
\begin{equation}
\Delta \Gamma=-\half  \left\langle \Delta \Psi\, \qbrs \left(\widetilde{\phix}_{i}\frac{\delta \Gamma}{\delta \meanf_{i}} +\widetilde{G}_{ij}\frac{\delta \Gamma}{\delta \prop_{ij}}+\widetilde{G}_{ijk\ldots}\frac{\delta \Gamma}{\delta G_{ijk\ldots}}+\ldots \right)^2 \right\rangle +O(\Delta^2)\label{eq:gfr}.
\end{equation}
where the quantities $\widetilde{G}_{ijk\ldots}$ are given by
\begin{subequations}
\bea
\widetilde{G}_{i}&=&\widetilde{\meanf}_i=\phix_i-\meanf_i, \\ 
\widetilde{G}_{ij}&=&
(\phix-\meanf)_i(\phix-\meanf)_j-G_{ij}, \\
\widetilde{G}_{ijk}
&=&(\phix-\meanf)_i(\phix-\meanf)_j(\phix-\meanf)_k-G_{ijk}-G_{ij}(\phix-\meanf)_k-G_{jk}(\phix-\meanf)_i-G_{ki}(\phix-\meanf)_j,\\
\widetilde{G}_{ijkl}&=&(\phix-\meanf)_i(\phix-\meanf)_j(\phix-\meanf)_k(\phix-\meanf)_l-G_{ijk}(\phix-\meanf)_l-G_{jkl}(\phix-\meanf)_i-G_{kli}(\phix-\meanf)_j-G_{lij}(\phix-\meanf)_k-\nonumber \\
&&\ \
G_{ij}G_{kl}-G_{ik}G_{jl}-G_{il}G_{kj}-\left\{G_{ij}\big[(\phix-\meanf)_k(\phix-\meanf)_l-G_{kl}\big]+\mbox{
5 permutations}\,\right\}-G_{ijkl},
\eea
\end{subequations}
etcetera.

\section{Gauge-fixing dependence of $\Phi$-derivable
approximations}\label{sec:four}

Our main interest is to study the gauge dependence of $\Phi$-derivable
approximations. As previously mentioned, they are
obtained after one truncates the \emph{skeleton-loop expansion} of the
2PI effective action at a certain order. For definiteness let us consider a truncation at $L$ loops, which translates into a truncation at $O(g^{2L-2})$ for the coupling constant $g$.  Then $\Gamma$ is split into two pieces 
\begin{equation}
\Gamma[\meanf,\prop]=\Gamma_0[\meanf,\prop]\,\mathbf{\big(}\mbox{of }
O(g^{2L-2})\mathbf{\big)}+\Gamma_1[\meanf,\prop]\,\mathbf{\big(}\mbox{of }
O(g^{2L})\mathbf{\big)},
\end{equation}
where the truncated part $\Gamma_0$ is used to generate \emph{approximate} mean fields $\meanf_{\ts{ap}}$ and two-point functions $\prop_{\ts{ap}}$  from the stationarity condition (\ref{stationaryrequirementtruncated}). \\

This splitting of the effective action can be performed directly on the result
(\ref{eq:fr}) for the variation $\Delta \Gamma$  under a shift of gauge
evaluated at the \emph{approximate} mean fields $\meanf_{\ts{ap}}$ and two-point
functions $G_{\ts{ap}}$ 
\begin{equation}
 \Delta \left(\Gamma_0+\Gamma_1\right)\left[ \meanf_{\ts{ap}}, \prop_{\ts{ap}}\right]=-\half \left\langle \Delta \Psi\, \qbrs \left( \frac{\delta \Gamma_1}{\delta \meanf_i}\Big|_{\meanf_{\ts{ap}},G_{\ts{ap}}}\widetilde{\phix}_{i,\,\ts{ap}} +\frac{\delta \Gamma_1}{\delta \prop_{jk}}\Big|_{\meanf_{\ts{ap}},G_{\ts{ap}}}\widetilde{G}_{jk,\,\ts{ap}} \right)^2 \right\rangle+O(\Delta^2),
\label{eq:frtruncated}
\end{equation} 
where we used the fact that $\meanf_{\ts{ap}}$ and $\prop_{\ts{ap}}$
correspond to the stationary point of $\Gamma_0$.\footnote{Since one works at
the stationary point of $\Gamma_0$ instead of the exact one obtained from
$\Gamma$ this implies immediately from Eq.~(\ref{trafos}) that expectation values $\langle \ldots\rangle$ are here evaluated at the values of the currents given by $J^{\prime}=-\delta \Gamma_1/\delta \meanf+2\meanf\,\delta \Gamma_1/\delta G$ and $K^{\prime}=-2\,\delta \Gamma_1/\delta G$. However, since $\Gamma_1\sim O(g^{2L})$, the expectation values are in first approximation equal to those obtained with vanishing currents.}\\

On one hand, Eq.~(\ref{eq:frtruncated}) implies that the \emph{truncated}
effective action $\Gamma_0$ evaluated at its corresponding physical mean
fields $\meanf_{\ts{ap}}$ and propagators $\prop_{\ts{ap}}$ is gauge independent up to the order
of truncation, i.e. $O(g^{2L-2})$. This is so since $\Gamma_1$ is of order
$O(g^{2L})$ and the r.h.s. of Eq.~(\ref{eq:frtruncated}) is of order
$O(\Gamma_1^2)$, so to first order $\Delta\Gamma_0 \approx -\Delta\Gamma_1
\approx O(g^{2L})$.\\

On the other hand, Eq.~(\ref{eq:frtruncated}) tells us that the
\emph{complete} action $\Gamma$ evaluated at the \emph{approximate} mean
fields and propagators obtained from $\Gamma_0$ is gauge-fixing independent up
to order $O(g^{4L})$, i.e. twice the order of $\Gamma_1$. This is a
consequence of having the square of the functional derivatives of $\Gamma_1$
in the r.h.s. of Eq.~(\ref{eq:frtruncated}) and can be understood with a
diagrammatic argument. To see that, first note\footnote{We take $\meanf=0$ for
simplicity.} that the diagrams in the loop expansion of the 2PI effective
action $\Gamma$ are
\emph{skeleton} diagrams, so without any self-energy insertions. The
approximate propagator $G_{\ts{ap}}$ obtained from truncating the skeleton
series of $\Gamma$ to $\Gamma_0$ at $L$ loops (or at $O(g^{2L-2})$) is the
solution of the variational condition (\ref{stationaryrequirement}), which can
be interpreted as a dressing of the bare propagator with all the self-energy
contributions that come from cutting one line in the 2PI diagrams of
$\Gamma_0$. Evaluating the effective action $\Gamma$ at $G_{\ts{ap}}$ entails
substituting this propagator in the diagrams of the \emph{skeleton-loop
expansion}. The outcome can be expanded perturbatively to compare directly
with the usual \emph{perturbative loop expansion} of the 1PI effective action.
One can check that both expansions match perfectly up to $2L$ loops, or
$O(g^{4L-2})$. They differ at $2L+1$ loops because, by construction, diagrams
that would result from dressing skeleton diagrams of $\Gamma_1$ with
self-energy contributions to $G_{\ts{ap}}$ coming also of $\Gamma_1$, do not
appear in the expansion of the skeleton series considered. However, they are
present in the perturbative loop expansion. The importance of the fact that
both expansions match up to $2L$ loops is that, since the perturbative loop expansion is gauge-invariant at every loop order, one can immediately conclude that so must be the skeleton-loop expansion of $\Gamma[G_{\ts{ap}}]$ up to $2L$ loops, or, in other words, up to $O(g^{4L})$. \\

In this manner, Eq. (\ref{eq:frtruncated}) shows that $\Phi$-derivable approximations, as truncations to the
2PI effective action, have a controlled gauge-fixing dependence, in the sense
that gauge dependent terms appear at higher orders. \\

\section{\bf Choice of gauge condition}\label{sec:five}
A large body of experience with gauge theory has led to the common view that
one should not tamper with gauge invariance. Yet, we explore here the
possibility of accepting a controlled amount of gauge dependence in the
computation of physical quantities. The question is then, what is a good
choice of gauge fixing? To be specific, consider the class of covariant
gauges described by $C^a=\partial^{\mu}A_{\mu}^a$. Then we have to decide on a
reasonable choice for the gauge parameter $\xi$. Evidently, $\xi$ should be
such that it does not upset the assumption that $\Gamma_1$ may be neglected
compared to $\Gamma_0$. \\

That such upset can happen is easier to see in the more
familiar perturbative case. There we have the loop expansion in terms of bare
propagators. Consider for simplicity diagrams without ghosts.  The gauge
propagators have a longitudinal part proportional to $\xi$, so in a given
diagram with $I$ internal lines we would have the factor $\xi^I$. In
terms of the number of three- and four-point vertices $V_3$ and $V_4$ and
the number of
loops $L$ it can be written
as $\xi^{2L-2+V_3/2}$. Together with the powers $g^{2L-2}$ in the bare coupling
constant $g$, the diagram has an overall factor $(g\xi)^{2L-2}\xi^{V_3/2}$.
Taking $\xi$ big enough, say $|\xi| > 1/g$, various terms belonging to
different orders of $g$ in the perturbation expansion would be shuffled. This
will evidently upset our ordering principle. \\

In a $\Phi$-derivable approximation, however, we consider the loop expansion
in terms of dressed propagators where their $\xi$-dependence is not clear
{\it a priori}.
For that one needs to find the stationary point of $\Gamma_0$. And this is
done after assuming that $\Gamma_1$ may be neglected
compared to $\Gamma_0$. Provided we had the explicit form of the dressed
propagator, an argument similar to the one above
would give the range of $\xi$ that is allowed without
 upsetting this assumption. Unfortunately, finding the dressed
propagators is in general a formidable task.
We nevertheless venture the following argument that a good
choice for $\xi$ is in the interval $(0,2)$. \\

Assume that the $\Phi$-derivable
approximation gives indeed an approximation to the path integral
\begin{equation}
Z=\int \mathcal{D}A\mathcal{D}c\mathcal{D}\bar{c}\,\, \mbox{exp} \left[
\frac{-i}{g^2}\int
d^4x\left\{\frac{1}{4}F^2_{g=1}+\bar{c}\partial^{\mu}(\partial_{\mu}-iA_{\mu})c+\frac{1}{2\xi}(\partial^{\mu}A_{\mu})^2\right\}
\right],
\end{equation}
where we have integrated out the $B$ field and rescaled $A \rightarrow A/g$,
$(c,\bar{c}) \rightarrow (c/\sqrt{g},\bar{c}/\sqrt{g})$. Then $g^2$ in the
above action is also the ordering parameter in the skeleton-loop expansion of
$\Phi$. If we do not want to upset this power counting, $\xi$ should be
treated of order one as $g^2 \rightarrow 0$. For finite $g^2$ it seems best to
choose $1/2\xi$ of the same magnitude as the other numerical coefficients in
the action, which are $1/4$ and $1/2$ for $F^2$, and 1 for the ghost terms. So
this suggests the choice $\xi$ in the range $1-2$. Saddle point arguments for
$g^2 \rightarrow 0$ are not upset by letting also $\xi \rightarrow 0$, so it is
reasonable to allow also values of $\xi \rightarrow 0$. On the other hand,
$\xi \gg 1$ would upset the longitudinal parts of the saddle point regions in
functional space (as for the perturbative case above). To allow a continuation
of the path integral to imaginary time, $\xi$ has to be positive. All these
arguments lead us to conclude that $\xi$ is best taken in the range $0-2$.



\section{Conclusions and comments}

In this paper, the gauge dependence of the 2PI effective action that
defines a $\Phi$-derivable approximation of a gauge theory has been determined.
To obtain it we used its definition as a Legendre transform of a generating
functional with bilocal sources and the BRS symmetry of the underlying
Yang-Mills action. As expected on general grounds, the result
shows that the 2PI effective action is gauge independent at its stationary
point. Furthermore, the result has been applied to study the gauge dependence
of $\Phi$-derivable approximations, defined as truncations of the
2PI effective action at a certain loop order. Even though correlation
functions derived within these approximation schemes are known not to fulfill
the Ward identities required by the gauge symmetry, it has been shown that the
truncated effective action defined at its stationary point has a controlled
gauge-fixing dependence, i.e. the explicit gauge dependent terms appear at
higher order. Furthermore, if one uses the stationary quantities of the truncated action to evaluate the \emph{complete} 2PI effective action, the gauge dependence appears then at twice the order of truncation. \\

These features might be interesting for the computation of thermodynamical
quantities derived from the 2PI effective action in gauge theories, such as
the pressure and entropy. The authors of~\cite{Blaizot:2000fc} have calculated
the entropy of the quark-gluon plasma with an approximate $\Phi$-derivable
approximation, but in order to achieve gauge independence, they had to
sacrifice the self-consistency guaranteed by working at the stationary point,
hence the word approximate. Their approach was nevertheless strongly motivated
from a quasiparticle picture of the quark-gluon plasma, which can be
used~\cite{Levai:1998yx} to describe the lattice
results~\cite{Boyd:1995zg}. In any case, our considerations
suggest that $\Phi$-derivable approximations may allow a systematic method
for computing thermodynamic functions without having to sacrifice its
remarkable properties. Gauge-fixing dependent artifacts would appear at high
orders, thus making the approximation controllable. \\

It might seem unsatisfactory that the gauge dependence is not completely removed in $\Phi$-derivable
approximations. Yet, we propose here to
accept a controlled amount of gauge
dependence in physical quantities. We argued that $\Phi$-derivable
approximations, provided they are indeed an approximation to the exact gauge
theory, implicitly restrict the choices of gauges available. This prevents the
high-order gauge artifacts to take arbitrary values that could render any
computed quantity physically meaningless. Both the fact that gauge dependent
terms appear at higher orders and that they are constrained by this
restriction makes the error introduced by breaking gauge invariance
controllable and indicate that
$\Phi$-derivable approximations may indeed give reasonable answers to physical
quantities. A detailed examination to quantify those gauge dependent
terms would involve solving a $\Phi$-derivable approximation for gauge
theories. So far, the complete solutions for QED and pure glue QCD even at
lowest order (2-loop) have not appeared in the literature.\\

We would like to note that in the derivation presented in this paper we did not discuss aspects
related to regularization and renormalization. This makes the calculations
heuristic in some respects. We compared the path integral with a skeleton expansion of the effective
action, neither of which has been clearly defined. To make the path integral
well defined a regularization is needed, preferably non-perturbative, and this
should be compatible with the BRS invariance used. With a
lattice regularization this is
non-trivial~\cite{Giusti:2001xf}. Another point is that the
regularization dependence needs to be removed, or at least shown to be
negligible (in the case of QED 'triviality' is expected to occur).
Renormalization is a non-trivial issue in $\Phi$-derivable
approximations~\cite{Braaten:2001vr}. A general renormalization
procedure, such as the one recently proposed by van Hees and
Knoll~\cite{vanHees:2001ik}, would be needed in order to have a well defined
path-integral. A detailed study of these issues in $\Phi$-derivable
approximations of gauge theories constitutes the subject of further
investigations.

\section*{Acknowledgements}
The authors wish to thank J.P. Blaizot, U. Reinosa, J. Berges, H. van Hees, E.
Mottola and C. van Weert for useful discussions and comments. The authors
would also like to thank the KITP in Santa Barbara for hospitality, and the
organizers of the workshop {\it QCD and Gauge Dynamics in the RHIC Era} for
the opportunity to participate. Research at the KITP was supported in part by
the National Science Foundation under Grant No. PHY99-07949. This research is
also supported by FOM.

\bibliography{references}
\end{document}